\documentclass[12pt,reqno]{amsart}

\usepackage{amsmath,amsthm,amscd,amsfonts,amssymb,graphicx,color}

\usepackage[bookmarksnumbered,colorlinks,plainpages]{hyperref}
\hypersetup{colorlinks=true,linkcolor=red,anchorcolor=green,citecolor=cyan,urlcolor=red,filecolor=magenta,pdftoolbar=true}

\usepackage{mathrsfs}

\newtheorem{theorem}{Theorem}[section]

\theoremstyle{definition}

\theoremstyle{remark}
\newtheorem{remark}[theorem]{Remark}

\numberwithin{equation}{section}

\begin{document}

\setcounter{page}{1}

\title[A phase space localization operator in negative binomial states]{
A phase space localization operator in negative binomial states}

\author[Zouha\"{\i}r Mouayn, Soumia Touhami ]{
 Zouha\"{\i}r Mouayn $^{1,2,3}$, Soumia Touhami $^4$}

\address{\footnotesize $^1$ Department of Mathematics, Faculty of Sciences and
Technics (M'Ghila), {\scriptsize Sultan Moulay Slimane University, B\'{e}ni Mellal, Morocco.}}
\address{\footnotesize $^2$ Institut des Hautes Études Scientifiques, Paris Saclay University
Le Bois-Marie, {\scriptsize 35 route de Chartres  CS 40001 91893 Bures-sur-Yvette, France.}}
\address{\footnotesize $^3$ Institut Henri Poincaré - UAR 839 Sorbonne University, / CNRS, {\scriptsize 11 rue Pierre et Marie Curie
75231 Paris Cedex 05 France.}}
\address{\footnotesize $^4$ Department of Mathematics, KTH Royal Institute
of Technology, {\scriptsize Stockholm, Sweden.}}
\email{\textcolor[rgb]{0.00,0.00,0.84}{mouayn@gmail.com  $^{1}$; touhami16soumia@gmail.com $^{2}$}}




\begin{abstract}
We are dealing with some spectral properties of a phase space localization
operator $P_{R}$ corresponding to the indicator function of a disk of radius 
$R<1.$ The localization procedure is achieved with respect to a set of
negative binomial states (NBS) labeled by points of the complex unit disk\ $%
\mathbb{D}$ and depending on a parameter $B>%
{\frac12}%
$ $.$ We derive a formula expressing $P_{R}$ as function of the
pseudo-harmonic oscillator whose potential function depends on $B$. The
phase space content outside the localization domain is estimated in terms of
the photon-counting probability distribution associated with the NBS. By
using the coherent states transform attached to NBS, we transfer the action
of the operator $P_{R}$ to a Bergman space\ $\mathcal{A}^{B}\left( \mathbb{D}%
\right) $ of analytic functions on\ $\mathbb{D}$ satisfying a growth
condition depending on $B$ and we explicitly give its integral kernel whose
limit\ as $R\rightarrow 1$ coincides with the reproducing kernel of \ $%
\mathcal{A}^{B}\left( \mathbb{D}\right) $. This leads to a natural
generalization of this Hilbert space with respect to the parameter $R$.

.
\end{abstract}

\maketitle


\section{\ \textbf{Introduction}}

The problem of localization in time and frequency has always been of serious
concern in modern physics because one of the major issues in applications is
to analyze signals on different time-frequency domains and therefore to
concentrate and localize signals on these domains. To be able to represent
the frequency behavior of a signal locally in time, one has to consider the
so called time-frequency localization operators. A variety of methods have
been invented to construct such class of operators \cite{W}.
Coherent states (CS) are the natural tool in constructing phase space
localization operators \ and have been extensively encountered in
theoretical physics, in quantum mechanics and in many different areas of
mathematical physics. Precisely, CS provide a close connection between
classical and quantum formalisms so as to play a central role in the semi
classical analysis. In general, they may be defined as an overcomplete
family of normalized ket vectors $|\zeta \rangle $ which are labeled by
points $\zeta $ of a phase-space domain $X$, belonging to a Hilbert space $%
\mathcal{H}$ that corresponds to a specific quantum model and provide $%
\mathcal{H}$ with a resolution of its identity operator as 
\begin{equation}
\label{r1}
\mathbf{1}_{\mathcal{H}}=\int_{X}|\zeta \rangle \langle \zeta |d\mu (\zeta ).
\end{equation}
with respect to a suitable integration measure $d\mu (\zeta )$ on $X$. These
states are constructed in different ways. For an overview of all aspects of
the theory of coherent states and their genesis, we refer to the \cite{V,TP}.

\smallskip

Equation $\eqref{r1} $ allows to implement a CS frame quantization \cite{G} of the set of parameters $\zeta \in X$ by associating to a
complex-valued function $\zeta \mapsto F(\zeta )$, satisfying appropriate
conditions, the following operator on $\mathcal{H}$ : 
\begin{equation}
\label{r2}
F(\zeta )\mapsto P_{F}:=\int\limits_{X}|\zeta \rangle \langle \zeta |F(\zeta
)d\mu (\zeta ).  
\end{equation}
If $F(\zeta )$ is semi-bounded real-valued function, the Friedrich extension 
\cite{RS} allows us to define $P_{F}$ as a self-adjoint operator.
In particular, when $F=\chi _{\Omega }$ is the indicator function for some
domain $\Omega $ in the phase space $X$, the resulting operator $P_{\chi
_{\Omega }}$ is called a localization operator.

\smallskip \smallskip

By using the CS of the harmonic oscillator, Daubechies \cite{Daub}
has discussed the localization operator $P_{\chi _{\Omega }}$ with $\Omega
\subset \mathbb{C}$ being a disk of radius $\rho >0$ by giving its
eigenfunctions in terms of Hermite polynomials, and by expressing its
discrete eigenvalues $\left\{ \lambda _{k}^{\rho }\right\} $ in term of
incomplete Gamma functions. She also has established the asymptotic behavior
of these eigenvalues for varying $k=0,1,2,...,$ and $\rho >0$, and has given
an estimate for the phase-content outside the localization domain $\Omega $.

\smallskip

In this paper, we deal with similar questions for the pseudo-harmonic
oscillator 

 \begin{equation}
      \label{r3}
H_{B}=\frac{1}{2}\left[ -\frac{d^{2}}{dx^{2}}+x^{2}+\frac{(2B-1)^{2}-\frac{1%
}{4}}{x^{2}}\right] +\left( 1-B\right) ,\text{ \ \ \ \ \ }2B>1  
\end{equation}
acting on the Hilbert space $L^{2}\left( \mathbb{R}_{+}\right) $, whose
importance consists in the fact that it is a solvable model and being, in a
certain sense, an intermediate potential between the three dimensional
harmonic oscillator potential and other anharmonic potentials such as
Poschl-Teller or Morse potential \cite{PD1, PD2}. The $L^{2}$
eigenfunctions (number states) of $H_{B}$, which here are denoted by the ket
vectors $\left\vert \ell _{j}^{B}\right\rangle ,$ \ may be superposed to
perform a set of coherent states within the so-called \textit{Hilbertian
probabilistic scheme} (see \cite{G}for the general theory) by
choosing a set of analytic coefficients $C_{j}^{B}\left( z\right) $ on the
complex unit disk $\mathbb{D=}\left\{ z\in \mathbb{C},\left\vert
z\right\vert <1\right\} $ such that the associated photon-counting
statistics follows a negative probability distribution. Such a CS are known
as the negative binomials states (NBS) \cite{Mou1}.\ One interest on
them is that they intermediate between pure coherent states and pure thermal
states \cite{GJT}and reduce to Susskind-Glogower phases states for
a particular limit of the parameter \cite{FS} Beside, such
coefficients $C_{j}^{B}\left( z\right) $, turn out to be basis elements of
the weighted Bergman space, here denoted $\mathcal{A}^{B}\left( \mathbb{D}%
\right) $, of analytic functions $g$ on $\mathbb{D}$, satisfying the growth
condition $\int_{\mathbb{D}}|g(z)|^{2}(1-\bar{z}z)^{2B-2}d\eta (z)<+\infty $%
, where $d\eta $ denotes the Lebesgue measure on $\mathbb{D}$.

Our aim is, firstly, to show that these NBS which are labeled by points of
the disk $\mathbb{D}$ can be retreived from the affine CS via the Cayley
transform. We also link them to the Landau problem in the Poincar\'{e} upper
half-plane. This connection may be exploited to generalize the obtained
results to higher hyperbolic Landau levels. Secondly, we proceed by a
quantization method based on these NBS in order to construct a phase space
localization operator $P_{R}$ corresponding to the disk $D_{R}=\left\{ z\in 
\mathbb{C},\left\vert z\right\vert <R\right\} $ with $R<1$, which stands for
the quantum counterpart of the classical observable defined as the indicator
function of the disk $D_{R}$. Precisely, we discuss some spectral properties
of the operator $P_{R}$ such as its eigenvalues and their associated
eigenfunctions in $L^{2}\left( \mathbb{R}_{+}\right) $. The expression of
these eigenvalues together with the discrete spectral resolution of $P_{R}$
amount to a formula expressing this operator as a function of the
Hamiltonian operator $H_{B}$ in $\eqref{r3} $. We also give an
estimate for the phase space content of $P_{R}$ outside the domain $D_{R}$
in terms of the photon-counting probability distribution associated with the
NBS. Moreover, this operator may be unitarly intertwined as $W\circ
P_{R}\circ W^{-1}$=$\widetilde{P}_{R}$ via the second Bargmann transform $W$
associated with the NBS. This allows us to obtain the integral kernel of $%
\widetilde{P}_{R}$ when acting on the space $\mathcal{A}^{B}\left( \mathbb{D}%
\right) $ by using calculations based on properties of some different
hypergeometric functions.

\smallskip

The paper is organized as follows. In section 2, we recall the affine
coherent states from which we derive the NBS.The connection with the Landau
problem on the Poincar\'{e} upper half-plane is also pointed out. Section 3
deals with the coherent states quantization method. In particular,
eigenvalues of the quantum counterpart with radial classical observables are
obtained.\ For the indicator function of the disk $D_{R}$ we also provide
these eigenvalues with a probabilistic interpretation and we discuss their
extensions to hyperbolic higher Landau levels. In section 4, we give an
estimate of the phase space content of the localization operator outside the
disk $D_{R}$ in terms of the photon-counting probablity distribution.
Section 5, we deal with the transfer of the localization operator to the
weigthed Bergman space $\mathcal{A}^{B}\left( \mathbb{D}\right) $ and to the
calculation of its integral kernel.

\section{ {\ \textbf{Negative binomial states and the $B$-weight Maass Laplacian}}}

\subsection{Affine coherent states}

We recall that the affine group is the set $\mathbf{G}=\mathbb{R}\times 
\mathbb{R}^{+}$, endowed with group law $\left( x,y\right) \cdot \left(
x^{\prime },y^{\prime }\right) =\left( x+yx^{\prime },yy^{\prime }\right) $. 
$\mathbf{G}$ is a locally compact group with the left Haar measure $d\nu
\left( x,y\right) =y^{-2}dxdy$. We shall consider one of the two
inequivalent infinite dimensional irreducible unitary representations of the
affine group $\mathbf{G}$, denoted $\pi _{+}$, realized on the Hilbert space 
$\mathcal{H}:=$ $L^{2}\left( \mathbb{R}^{+},\xi ^{-1}d\xi \right) $ as 
\begin{equation}
\label{r4}
\pi _{+}\left( x,y\right) \left[ \varphi \right] \left( \xi \right) :=e^{%
\frac{1}{2}ix\xi }\varphi \left( y\xi \right) ,\qquad \varphi \in \mathcal{\
H},\quad \xi >0\text{.}  
\end{equation}
This representation is square integrable since it is easy to find a vector $%
\phi _{0}\in \mathcal{H}$ such that the function $\left( x,y\right) \mapsto
\left\langle \pi _{+}\left( x,y\right) \left[ \phi _{0}\right] ,\phi
_{0}\right\rangle _{\mathcal{H}}$ belongs to $L^{2}\left( \mathbf{G},d\nu
\right) $. This condition can also be expressed by saying that the
self-adjoint operator $\delta :\mathcal{H\rightarrow H}$ defined as $\delta %
\left[ \varphi \right] (\xi )=\xi ^{-\frac{1}{2}}\varphi \left( \xi \right) $
gives 
\begin{equation}
\label{r5}
\int\limits_{\mathbf{G}}\left\langle \varphi _{1},\pi _{+}\left( x,y\right) %
\left[ \psi _{1}\right] \right\rangle \left\langle \pi _{+}\left( x,y\right) %
\left[ \varphi _{2}\right] ,\psi _{2}\right\rangle d\nu \left( x,y\right)
=\left\langle \varphi _{1},\varphi _{2}\right\rangle \left\langle \delta ^{%
\frac{1}{2}}\left[ \varphi _{1}\right] ,\delta ^{\frac{1}{2}}\left[ \varphi
_{2}\right] \right\rangle  
\end{equation}
for all $\psi _{1},\psi _{2},\varphi _{1},\varphi _{2}\in \mathcal{H}$. The
operator $\delta $ is unbounded because $\mathbf{G}$ is not unimodular \cite{DMo}.

\smallskip

Keeping the condition $2B>1,$ we consider a set of CS labeled by elements $%
(x,y)\in \mathbf{G}$, which are obtained by acting, via the representation
operator $\pi _{+}\left( x,y\right) $, on the admissible vector 
\begin{equation}
\label{r6}
\phi _{B}\left( \xi \right) :=\frac{1}{\sqrt{2B}}\xi ^{B}e^{-\frac{1}{2}\xi
},\text{ \ \ }\xi >0.  
\end{equation}
Precisely, 
\begin{equation}
\label{r7}
\left\vert \tau _{(x,y),B}\right\rangle :=\pi _{+}\left( x,y\right) \left[
\phi _{B}\right]  
\end{equation}
and satisfy the resolution of the identity operator 
\begin{equation}
\label{r8}
\mathbf{1}_{\mathcal{H}}=c_{B}\int\limits_{\mathbf{G}}d\mu \left( x,y\right)
\left\vert \tau _{(x,y),B}\right\rangle \left\langle \tau
_{(x,y),B}\right\vert 
\end{equation}%
where $c_{B}:=2B-1$ and the Dirac's bra-ket notation $|\Phi \rangle \langle
\Phi |$ means the rank-one operator $\phi \longmapsto \langle \Phi ,\phi
\rangle _{\mathcal{H}}.\Phi $ with $\Phi ,\phi \in \mathcal{H}$. In the $\xi 
$-coordinate, wavefunctions of CS defined by Eq. $\eqref{r7} $ read 
\begin{equation}
\label{r9}
\left\langle \xi \right\vert \tau _{(x,y),B}\rangle =\frac{1}{\sqrt{2B}}%
\left( \xi y\right) ^{B}e^{-\frac{1}{2}\xi \left( y-ix\right) }\text{, }%
\qquad \xi >0,  
\end{equation}
and are known as the affine CS \cite{AK}.

\subsection{Connection with the $B$-weight Maass Laplacian}

To describe the connection of CS $\eqref{r9} $ with the lowest
hyperbolic Landau level, we may first identify the affine group $\mathbf{G}$
with the Poincar\'{e} upper half-plane $\mathbb{H}^{2}=\left\{ x+iy,x\in 
\mathbb{R},y>0\right\} $. Then, to these CS we may attach, as usual, the CS
transform $\mathcal{B}_{0}:\mathcal{H}\rightarrow L^{2}\left( \mathbb{H}%
^{2},d\nu \right) $ defined by \cite{Mou2}: 
\begin{equation}
\label{r10}
\mathcal{B}_{0}[\phi ]\left( x,y\right) =\sqrt{c_{B}}\int\limits_{0}^{+%
\infty }\overline{\left\langle \xi \right\vert \tau _{(x,y),B}\rangle }\phi
(\xi )\xi ^{-1}d\xi   
\end{equation}
whose range is the eigenspace of the $B$-weight Maass Laplacian 
\begin{equation}
\Delta _{B}=y^{2}\left( \partial _{x}^{2}+\partial _{y}^{2}\right)
-2iBy\partial _{x},  
\end{equation}
associated with the eigenvalue
\begin{equation}
\label{r11}
\epsilon _{m}^{B}=(B-m)\left( 1-B+m\right) ,\quad m=0,1,...,\left\lfloor B-%
{\frac12}%
\right\rfloor ,  
\end{equation}
where $\left\lfloor a\right\rfloor $ denotes the greatest integer not
exceeding $a.$ We precisely have 
\begin{equation}
\label{r12}
\mathcal{B}_{0}[\mathcal{H}]\equiv \left\{ f\in L^{2}\left( \mathbb{H}%
^{2},d\nu \right) ,\;\Delta _{B}f=\epsilon _{0}^{B}f\right\} .  
\end{equation}
The operator $\Delta _{B}$ also stands (in suitable units and up to an
additive constant) for the Schr\"{o}dinger operator describing the dynamics
of a charged particle moving on $\mathbb{H}^{2}$ under the action of a
magnetic field of strength proportional to $B.$ This is an elliptic densely
defined operator on the Hilbert space $L^{2}(\mathbb{H}^{2},d\nu )$, with a
unique self-adjoint realization also denoted by $\Delta _{B}$. Its spectrum
consists of two parts:\textit{\ }a continuous part $\left[ 
{\frac14}%
\theta ,+\infty \right[ $, corresponding to scattering states and the finite
number of eigenvalues $\epsilon _{m}^{B}$ each one with infinite
degeneracy,\ called hyperbolic Landau levels . Finally, the reproducing
kernel of the Hilbert space $\mathcal{B}_{0}[\mathcal{H}]$ can be obtained
from the overlapping function $\langle \tau _{w,B},\tau _{\zeta ,B}\rangle _{%
\mathcal{H}}$ between two CS as 
\begin{equation}
\label{r13}
K_{0}^{B}\left( w,\zeta \right) =\left( \frac{\left\vert w-\bar{\zeta}\right\vert ^{2}}{4\rm{Im}w\rm{Im}\zeta }\right) ^{-B}\left( \frac{\zeta-\bar{w}}{w-\bar{\zeta}}\right) ^{B}{},\text{ \ }w,\zeta \in \mathbb{H}^{2}.
\end{equation}

\subsection{Negative binomial states}

\smallskip We can write a version of these CS as states labeled by points $z$
of the unit disk $\mathbb{D}$ by using the inverse Cayley transform $%
\mathcal{C}^{-1}:\mathbb{D}\rightarrow \mathbf{G}$ given by 
\begin{equation}
\label{r14}
\mathcal{C}^{-1}\left( z\right) =\left( -2\frac{\rm{Im}z}{\left\vert
1-z\right\vert ^{2}},\frac{1-\left\vert z\right\vert ^{2}}{\left\vert
1-z\right\vert ^{2}}\right) ,\text{ \ }z\in \mathbb{D}.  
\end{equation}
Indeed, we may still define this version as states in $\mathcal{H}$ as%
\begin{equation}
\label{r15}
\kappa _{z,B}:=\left( \frac{1-\bar{z}}{1-z}\right) ^{B}\pi _{+}\left( 
\mathcal{C}^{-1}\left( z\right) \right) \left[ \phi _{B}\right] . 
\end{equation}
Direct calculations lead to their wave functions in the $\xi $-coordinate as 
\begin{equation}
\label{r16}
\left\langle \xi \right\vert \kappa _{z,B}\rangle =\frac{1}{\sqrt{\Gamma
\left( 2B\right) }}\left( \frac{\left( 1-z\bar{z}\right) }{\left( 1-z\right)
^{2}}\xi \right) ^{B}\exp \left( -\frac{1}{2}\left( \frac{1+z}{1-z}\right)
\xi \right) ,\qquad \xi >0.  
\end{equation}
The latter ones may slightly be modified in order to perform them as vectors
of $L^{2}(\mathbb{R}_{+},d\xi )$ labeled by points of the disk $\mathbb{D}$
as 
\begin{equation}
\label{r17}
\left\langle \xi \right\vert \widetilde{\kappa }_{z,B}\rangle :=\sqrt{\frac{2%
}{\xi }}\,\langle \xi ^{2}|\kappa _{z,B}\rangle  
\end{equation}
These states obey the normalization condition $\langle \widetilde{\kappa }%
_{z,B},\widetilde{\kappa }_{z,B}\rangle _{L^{2}(\mathbb{R}_{+})}=1$ and
satisfy the resolution of the identity operator as 
\begin{equation}
\label{r18}
1_{L^{2}(\mathbb{R}_{+})}=\int\limits_{\mathbb{D}}\left\vert \widetilde{%
\kappa }_{z,B}\right\rangle \left\langle \widetilde{\kappa }%
_{z,B}\right\vert d\eta _{B}(z),  
\end{equation}
with respect to the measure 
\begin{equation}
\label{r19}
d\eta _{B}(z):=\frac{(2B-1)}{\pi \left( 1-z\overline{z}\right) ^{2}}d\eta
(z).  
\end{equation}

Now, as for the canonical CS of the harmonic oscillator, we seek for a
number states expansion of the CS $\left\vert \widetilde{\kappa }%
_{z,B}\right\rangle $ in terms of the analytic coefficients 
\begin{equation}
\label{r20}
C_{j}^{B}(z):=(2B-1)^{1/2}\sqrt{\frac{\Gamma (2B+j)}{j!\Gamma (2B)}}z^{j},%
\text{ \ \ }j=0,1,2,...,  
\end{equation}
which constitute an orthonormal basis of the weighted Bergman space $%
\mathcal{A}^{B}(\mathbb{D})$. For that, we start from the expression of the
CS, $\left\vert \widetilde{\kappa }_{z,B}\right\rangle $ as given by $\eqref{r17}-\eqref{r16}$  and we make use of the generating function
for the Laguerre polynomials (\cite{MaOb}, p.239): 
\begin{equation}
\label{r21}
\sum_{j=0}^{+\infty }t^{j}L_{j}^{(\alpha )}(x)=\frac{1}{(1-t)^{\alpha +1}}%
\exp \left( \frac{-t}{1-t}x\right) ,\ \ \alpha >-1.  
\end{equation}
We obtain, after some calculations, the series expansion 
\begin{equation}
\label{r22}
\langle \xi \left\vert \widetilde{\kappa }_{z,B}\right\rangle =\left( \frac{%
2B-1}{(1-z\bar{z})^{2B}}\right) ^{-1/2}\sum_{j=0}^{+\infty }C_{j}^{B}(z)\ell
_{j}^{B}(\xi )  
\end{equation}
where $\ell _{j}^{B}(\xi )$ are the Laguerre functions 
\begin{equation}
\label{r23}
\ell _{j}^{B}(\xi ):=\left( \frac{2j!}{\Gamma (2B+j)}\right) ^{1/2}\xi ^{2B-%
\frac{1}{2}}e^{-\frac{1}{2}\xi ^{2}}L_{j}^{(2B-1)}(\xi ^{2}),\text{ \ \ }%
j=0,1,2,..., 
\end{equation}
which are known to constitute a complete orthonormal system in $L^{2}(%
\mathbb{R}_{+},d\xi )$.

\smallskip

 \begin{remark}
    
The CS $\eqref{r16} $ also coincide with those
constructed by Molanar \textit{et al} \cite{BMMB} for the Morse
potential \cite{MP} by an algebraic way based on supersymetry and
shape invariance properties where the shape parameter may be taken as our $%
B>0$. For this potential, they first were were introduced by Nieto \textit{%
et al} \cite{NMSL} as generalized minimal uncertainty
states.
\end{remark}

\section{{\ \textbf{Quantization via $\ $NBS $\left\vert \widetilde{\protect\kappa }%
_{z,B}\right\rangle $}}}

The resolution of the identity $\eqref{r18}$ allows to implement a
CS or frame quantization \cite{TP} of the set of parameters $%
\mathbb{D}$ by associating to a function $\mathbb{D}\ni z\mapsto F(z,%
\overline{z})\in \mathbb{C}$ that satisfies appropriate conditions, the
following operator in $L^{2}(\mathbb{R}_{+},d\xi ):$ 
\begin{equation}
\label{r24}
F\mapsto \wp _{F}^{B}:=\int\limits_{\mathbb{D}}\left\vert \widetilde{\kappa }%
_{z,B}\right\rangle \left\langle \widetilde{\kappa }_{z,B}\right\vert F(z,%
\overline{z})\frac{(2B-1)}{\pi (1-z\bar{z})^{2}}d\eta (z).  
\end{equation}
The Friedrich extension \cite{RS} allows to define $\wp _{F}^{B}$
as a self-adjoint operator if $F$ is a semi-bounded real-valued function.%
\newline

\bigskip \smallskip

In order to proceed with the quantization through the CS $\left\vert 
\widetilde{\kappa }_{z,B}\right\rangle $ along the linear map $\eqref{r24} $, we may substitute the expression $\eqref{r22}$ into
the integral form in $\eqref{r24} .$ This gives the expression%
\begin{equation}
\label{r25}
\wp _{F}^{B}=\sum_{j,k=0}^{+\infty }\sqrt{\frac{\Gamma (2B+j)}{\pi j!\Gamma
(2B)}}\sqrt{\frac{\Gamma (2B+k)}{\pi k!\Gamma (2B)}}\left[
(2B-1)\int\limits_{\mathbb{D}}z^{k}\bar{z}^{j}(1-z\bar{z})^{2B-2}F(z,\bar{z}%
)d\eta (z)\right] \left\vert \ell _{j}^{B}\right\rangle \left\langle \ell
_{k}^{B}\right\vert  
\end{equation}
which represents its disrete spectral resolution\textit{\ }%
\begin{equation}
\label{r26}
\wp _{F}^{B}=\sum_{j,k=0}^{+\infty }\left[ \gamma _{F}^{B}\right]
_{j,k}\left\vert \ell _{k}^{B}\rangle \langle \ell _{j}^{B}\right\vert 
\end{equation}
where the matrix elements are (at least formally) given by\textit{\ }%
\begin{equation}
\label{r27}
\left[ \gamma _{F}^{B}\right] _{j,k}=\frac{1}{\pi \Gamma (2B-1)}\left( \frac{%
\Gamma (2B+j)\Gamma (2B+k)}{j!k!}\right) ^{1/2}\int\limits_{\mathbb{D}}\bar{z%
}^{j}z^{k}(1-z\bar{z})^{2B-2}F(z,\bar{z})d\eta (z)  
\end{equation}
and\ $\left\{ \ell _{j}^{B}\right\} $ is the orthonormal basis of $L^{2}(%
\mathbb{R}_{+},d\xi ),$\ which is given in $\eqref{r23}.$

\subsection{Radial classical observables}

For a radial weight function $F$, the above discrete spectral resolution of $%
\wp _{F}^{B}$ leads to more precise expressions for its eigenvalues. Indeed,
by setting $F(z,\bar{z})=F(r^{2})$, $r=\left\vert z\right\vert $ and using
polar coordinates in the expression $\eqref{r27} $ of matrix elements,
we get that 
\begin{equation}
\label{r28}
\left[ \gamma _{F}^{B}\right] _{j,k}=\frac{1}{\pi \Gamma (2B-1)}\left( \frac{%
\Gamma (2B+j)\Gamma (2B+k)}{j!k!}\right) ^{1/2}\int\limits_{0}^{2\pi
}\int\limits_{0}^{1}r^{k}r^{j}e^{ik\theta }e^{-ij\theta
}(1-r^{2})^{2B-2}F(r^{2})rdrd\theta .  
\end{equation}
By the fact that 
\begin{equation}
\label{r29}
\int\limits_{0}^{2\pi }e^{i(k-j)\theta }d\theta =2\pi \delta _{k,j},\text{ \ 
}j,k=0,1,2,...,  
\end{equation}
on can easily see that only the case $j=k$ produces a non zero matrix
element as 
\begin{equation}
\label{r30}
\left[ \gamma _{F}^{B}\right] _{j,j}=\frac{\Gamma (2B+j)}{\Gamma
(2B-1)\Gamma (j+1)}\int_{0}^{1}\rho ^{j}(1-\rho )^{2B-2}F(\rho )d\rho . 
\end{equation}
By writing the prefactor as $\left( \mathcal{B}(j+1,2B-1)\right) ^{-1}$, we
obtain the expression of the $\lambda _{j}^{B,F}$ as

\begin{equation}
\label{r31}
\lambda _{j}^{B,F}=\frac{1}{\mathcal{B}(j+1,2B-1)}\int\limits_{0}^{1}\rho
^{j}(1-\rho )^{2B-2}F(\rho )d\rho ,  
\end{equation}
where $\mathcal{B}(a,b)$ denotes the Beta function with $a,b>0$. The
operator $\wp _{F}^{B}$ has the following discrete spectral resolution with
respect to the orthonormal basis $\{\ell _{j}^{B}\}$ as 
\begin{equation}
\label{r32}
\wp _{F}^{B}=\sum_{j=0}^{+\infty }\lambda _{j}^{B,F}\left\vert \ell
_{j}^{B}\right\rangle \left\langle \ell _{j}^{B}\right\vert ,  
\end{equation}
and it's not difficult to check that 
\begin{equation}
\label{r33}
\wp _{F}^{B}\left[ \ell _{j}^{B}\right] =\lambda _{j}^{B,F}\ell _{j}^{B}. 
\end{equation}
Note that $\eqref{r33} $ means that the operators $\wp _{F}^{B}$ and $%
H_{B}$ have $\{\ell _{j}^{B}\}$ as a commun set of eigenfunctions.

\smallskip

In particular, we here consider the disk $D_{R}:=\{z\in \mathbb{C},\ |z|<R\}$
with $0<R<1$ and we choose as classical observable the indicator function of
this domain. By putting $F(r^{2})=1$ if $r<R$ and $F(r^{2})=0$ if $r\geq R$,
the formula $\eqref{r31} $ takes the form 
\begin{equation}
\label{r34}
\lambda _{j}^{B,R}=\frac{1}{\mathcal{B}(j+1,2B-1)}\int\limits_{0}^{R^{2}}%
\rho ^{j}(1-\rho )^{2B-2}d\rho =\mathcal{I}_{R^{2}}(j+1,2B-1)  
\end{equation}
where 
\begin{equation}
\label{r35}
\mathcal{I}_{x}(a,b)=\frac{1}{\mathcal{B}(a,b)}\int%
\limits_{0}^{x}t^{a-1}(1-t)^{b-1}dt,\quad 0<x<1,\text{ }  
\end{equation}
is the regularized incomplete Beta function. In view of $\eqref{r32}$, the discrete spectral resolution of $\wp _{\digamma }^{B}$
reads 
\begin{equation}
\label{r36}
\wp _{R}^{B}=\sum_{j=0}^{+\infty }\mathcal{I}_{R^{2}}(j+1,2B-1)\left\vert
\ell _{j}^{B}\right\rangle \left\langle \ell _{j}^{B}\right\vert . 
\end{equation}
By another side, the vector basis $\ell _{j}^{B}$ are eigenfunctions of $%
H_{B}$ while acting on $L^{2}\left( \mathbb{R}_{+},d\xi \right) $.
Precisely, 
\begin{equation}
\label{r37}
H_{B}\left[ \ell _{j}^{B}\right] =(j+1)\ell _{j}^{B},\ j=0,1,2,...\text{ .} 
\end{equation}
Therefore, we may write $\wp _{R}^{B}$ as a function of $H_{B}$ as 
\begin{equation}
\label{r38}
\wp _{R}^{B}=\mathcal{I}_{R^{2}}(j+1,2B-1).  
\end{equation}

\subsection{A probabilistic representation for eigenvalues $\protect\lambda %
_{j}^{B,R}$}

We note that the eigenvalues $\eqref{r34} $ may also be written as 
\begin{equation}
\label{r39}
\lambda _{j}^{B,R}=\parallel C_{j}^{B}1_{\mathbb{D}_{R}}\parallel _{L^{2}(%
\mathbb{D},(1-z\bar{z})^{2B-2}d\eta )}^{2}=\int_{0}^{R^{2}}\mathfrak{g}%
_{j,B}(\rho )d\rho  
\end{equation}
where 
\begin{equation}
\label{r39}
\mathfrak{g}_{j,B}(\rho )=\frac{\Gamma (2B+j)}{\Gamma (2B-1)\Gamma (j+1)}%
(1-\rho )^{2B-2}\rho ^{j},\quad 0\leq \rho <1  
\end{equation}
which turns out to be the identity function of the Beta distribution $%
\mathcal{Y}_{j,B}^{\left( 0\right) }\sim \mathcal{B}e(j+1,2B-1)$ whose
characteristic function is known to be given by the confluent hypergeometric
series as $u\mapsto {\ }_{1}F_{1}\left( j+1,2B+1;iu\right) $ with $i^{2}=-1.$%
Therefore, 
\begin{equation}
\label{r40}
\lambda _{j}^{B,R}=\Pr (\mathcal{Y}_{j,B}^{\left( 0\right) }\leq R^{2}) 
\end{equation}
would provide us with a probabilistic representation of these eigenvalues.

\smallskip

For higher hyperbolic Landau levels$,$ the generalized form of the density
function $\eqref{r39}$ is given by 
\begin{equation}
\label{r41}
\mathfrak{g}_{B,j}^{\left( m\right) }(\rho ):=\left( 2B-2m-1\right) \frac{%
\left( m\wedge j\right) !}{\left( m\vee j\right) !}\frac{\Gamma (2B-2m+m\vee
j)}{\Gamma (2B-2m+m\wedge j)}(1-\rho )^{2B-2m-2}\rho ^{\left\vert
m-j\right\vert }  
\end{equation}
\begin{equation*}
\times \left( P_{m\wedge j}^{\left( \left\vert m-j\right\vert ,2\left(
B-m\right) -1\right) }\left( 1-2\rho \right) \right) ^{2}
\end{equation*}%
where $P_{k}^{\left( \alpha ,\beta \right) }$ $\left( .\right) $ $\ $is a
Jacobi polynomial \cite{MaOb} $\ $and $m=0,1,...,\left\lfloor B-%
{\frac12}%
\right\rfloor .$ Let us denote by $\mathcal{Y}_{j,B}^{\left( m\right) }$ the
random variable having $\rho \mapsto \mathfrak{g}_{j,B}^{\left( m\right)
}(\rho )$ as its density, then 
\begin{equation}
\label{r42}
\lambda _{j}^{B,R,m}:=\Pr \left( \mathcal{Y}_{j,B}^{\left( m\right) }\leq
R^{2}\right) =\int\limits_{0}^{R^{2}}\mathfrak{g}_{j,B}^{\left( m\right)
}(\rho )d\rho   
\end{equation}
would provide us with the probabilistic representation of eigenvalues $%
\lambda _{j}^{B,R,m}$ of the restricted operator $\mathfrak{K}_{B,m}\mid
_{D_{R}}$to the disk $D_{R},$ where $\mathfrak{K}_{B,m}$ is the projection
operator onto the eigenspace
\begin{equation}
\label{r43}
\mathcal{E}_{B,m}\left( \mathbb{D}\right) =\left\{ f\in L^{2}\left( \mathbb{D%
},\left( 1-z\overline{z}\right) ^{2B-2}d\eta \left( z\right) \right) ,%
\widetilde{\Delta }_{B}f=\sigma _{B,m}f\right\}   
\end{equation}
of the $B$-weight Maass Laplacian 
\begin{equation}
\label{r44}
\widetilde{\Delta }_{B}=-4\left( 1-z\overline{z}\right) \left( \left( 1-z%
\overline{z}\right) \frac{\partial ^{2}}{\partial z\partial \overline{z}}-2B%
\overline{z}\frac{\partial }{\partial \overline{z}}\right) ,  
\end{equation}
associated with the hyperbolic Landau level%
\begin{equation}
\label{r45}
\sigma _{B,m}=4m\left( 2B-m-1\right) ,\text{ \ \ }m=0,1,...,\left\lfloor B-%
{\frac12}%
\right\rfloor .  
\end{equation}
So that the integral kernel of operator $\mathfrak{K}_{B,m}$ is the
reproducing kernel 
\begin{equation}
\label{r46}
K_{B,m}\left( z,w\right) =\pi \left( 2B-2m-1\right) \left( 1-z\overline{w}%
\right) ^{-2B}\left( \frac{\left\vert 1-z\overline{w}\right\vert ^{2}}{%
\left( 1-z\overline{z}\right) \left( 1-w\overline{w}\right) }\right) ^{m} 
\end{equation}
\begin{equation*}
\times P_{m}^{\left( 0,2\left( B-m\right) -1\right) }\left( 2\frac{\left( 1-z%
\overline{z}\right) \left( 1-w\overline{w}\right) }{\left\vert 1-z\overline{w%
}\right\vert ^{2}}-1\right) 
\end{equation*}
of the eigenspace $\eqref{r46}.$ Note that for $2B>1$ and $m=0$ this
eigenspace reduces to the weighted Bergman space of analytic functions $g$
on $\mathbb{D}$, satisfying the growth condition $\int_{\mathbb{D}%
}|g(z)|^{2}(1-\bar{z}z)^{2B-2}d\eta (z)<+\infty $. In other words, $\mathcal{%
E}_{B,0}\left( \mathbb{D}\right) \equiv \mathcal{A}^{B}\left( \mathbb{D}%
\right) .$ This remark makes possible to extend our analysis to the above
higher hyperbolic Landau levels $\sigma _{B,m}.$ In this respect the
connection with the results in \cite{CDM} may be useful.

\section{{\ \textbf{Phase space content of $\wp _{R}^{B}$ outside $D_{R}$}}}

Note that the phase space cutoff by the operator $\wp _{R}^{B}$ is not sharp
in the sense that it will have some phase space content outside the
localization domain $D_{R}$. This is illustrated by the fact that at least
for some coherence point $z_{0}\in \mathbb{D}\setminus D_{R}$ we have 
\begin{equation}
\label{r47}
\langle \widetilde{\kappa }_{z_{0},B}\left\vert \wp _{R}^{B}[f]\right\rangle
_{L^{2}(\mathbb{R}_{+})}\neq 0,\quad f\in L^{2}(\mathbb{R}_{+}).  
\end{equation}
Precisely, we have the following estimate involving the photon counting
statistics which obey the negative binomial probability distribution $%
\mathcal{X}$ $\sim \mathcal{NB}(2B,z_{0}\bar{z}_{0})$\textit{\ }with\textit{%
\ }parameters $2B$\ and $z_{0}\bar{z}_{0}$ 
\begin{equation}
\label{r48}
\left\vert \langle \widetilde{\kappa }_{z_{0},B}\left\vert \wp
_{R}^{B}[f]\right\rangle _{L^{2}\left( \mathbb{R}_{+}\right) }\right\vert
\leq \sqrt{\mathbb{E}\left( \left( \mathcal{I}_{R^{2}}\left( \mathcal{X+}%
1,2B-1\right) \right) ^{2}\right) }\Vert f\Vert _{L^{2}\left( \mathbb{R}%
_{+}\right) } . 
\end{equation}
To prove $\eqref{r48}$ we start by replacing in the scalar product 
$\eqref{r47}$ the operator $\wp _{R}^{B}$ by its discrete spectral
resolution $\eqref{r36}$ as 
\begin{equation}
\label{r49}
\left\langle \widetilde{\kappa }_{z_{0},B},\left( \sum_{j=0}^{+\infty
}\lambda _{j}^{B,R}\left\vert \ell _{j}^{B}\right\rangle \left\langle \ell
_{j}^{B}\right\vert \right) [f]\right\rangle _{L^{2}(\mathbb{R}%
_{+})}=\sum_{j=0}^{+\infty }\lambda _{j}^{B,R}\left\langle \widetilde{\kappa 
}_{z_{0},B}|\ell _{j}^{B}\right\rangle _{L^{2}\left( \mathbb{R}_{+}\right)
}\left\langle \ell _{j}^{B}|f\right\rangle _{L^{2}\left( \mathbb{R}%
_{+}\right) }.  
\end{equation}
Recalling the number states expansion $\eqref{r22} $ of $\left\vert 
\widetilde{\kappa }_{z,B}\right\rangle $, we may write 
\begin{equation}
\label{r50}
\left\langle \widetilde{\kappa }_{z_{0},B}|\ell _{j}^{B}\right\rangle
_{L^{2}\left( \mathbb{R}_{+}\right) }=(1-z_{0}\bar{z}_{0})^{B}\sqrt{\frac{%
\Gamma (2B+j)}{j!\Gamma (2B)}}z_{0}^{j}.  
\end{equation}
Therefore $\eqref{r49} $ reads 
\begin{equation}
\label{r51}
\langle \widetilde{\kappa }_{z_{0},B}\left\vert \wp _{R}^{B}[f]\right\rangle
_{L^{2}\left( \mathbb{R}_{+}\right) }=(1-z_{0}\bar{z_{0}})^{B}\sum_{j=0}^{+%
\infty }\lambda _{j}^{B,R}\left( \frac{\Gamma (2B+j)}{j!\Gamma (2B)}\right)
^{1/2}\left\langle \ell _{j}^{B}|f\right\rangle _{L^{2}\left( \mathbb{R}%
_{+}\right) }z_{0}^{j}.  
\end{equation}
\begin{equation}
\label{r52}
=(1-z_{0}\bar{z_{0}})^{B}\left\langle \sum_{j=0}^{+\infty }\lambda
_{j}^{B,R}\left( \frac{\Gamma (2B+j)}{j!\Gamma (2B)}\right)
^{1/2}z_{0}^{j}\ell _{j}^{B}|f\right\rangle _{L^{2}(\mathbb{R}_{+})}. 
\end{equation}
Setting 
\begin{equation}
\label{r53}
\vartheta :=\sum_{j=0}^{+\infty }\lambda _{j}^{B,R}\left( \frac{\Gamma (2B+j)%
}{j!\Gamma (2B)}\right) ^{1/2}z_{0}^{j}\ell _{j}^{B},  
\end{equation}
and using the Cauchy-Bunyakovsky-Schwarz inequality 
\begin{equation}
\label{r54}
\left\vert \langle \widetilde{\kappa }_{z_{0},B}|\wp _{R}^{B}[f]\rangle
_{L^{2}\left( \mathbb{R}_{+}\right) }\right\vert \leq |1-z_{0}\bar{z_{0}}%
|^{B}\Vert \vartheta \Vert _{L^{2}(\mathbb{R}_{+})}\left\Vert f\right\Vert
_{L^{2}\left( \mathbb{R}_{+}\right) }.
\end{equation}
Now, the $L^{2}$ norm of $\vartheta $ may be written 
\begin{equation}
\label{r54}
|1-z_{0}\bar{z_{0}}|^{B}\Vert \vartheta \Vert _{L^{2}(\mathbb{R}%
_{+})}=\left( \sum_{j=0}^{+\infty }\left( \lambda _{j}^{B,R}\right) ^{2}%
\left[ \frac{\Gamma (2B+j)}{j!\Gamma (2B)}(z_{0}\bar{z_{0}})^{j}(1-z_{0}\bar{%
z_{0}})^{2B}\right] \right) ^{1/2}.
\end{equation}
By recognizing in $\eqref{r54} $ the probability distribution of $%
\mathcal{X}\sim \mathcal{N}\mathcal{B}(2B,z_{0}\overline{z_{0}})$ : 
\begin{equation}
\label{r55}
\Pr \left( \mathcal{X}=j\right) =\frac{\Gamma (2B+j)}{j!\Gamma (2B)}\left(
z_{0}\overline{z_{0}}\right) ^{j}(1-z_{0}\overline{z_{0}})^{2B},\ \
j=0,1,2,\cdots ,  
\end{equation}
and using the expression of the eigenvalues $\lambda _{j}^{B,R}$, Eq.$\eqref{r54} $ takes the form 
\begin{equation}
\label{r56}
|1-z_{0}\bar{z_{0}}|^{B}\Vert \vartheta \Vert _{L^{2}(\mathbb{R}%
_{+})}=\left( \sum_{j=0}^{+\infty }\left( \mathcal{I}_{R^{2}}(j+1,2B-1)%
\right) ^{2}\Pr \left( \mathcal{X}=j\right) \right) ^{1/2}. 
\end{equation}
Finally, the right hand side of $\eqref{r56} $  can be viewed as the
square root of an expectation value as $\mathbb{E}\left( \left( \mathcal{I}%
_{R^{2}}(\mathcal{X}+1,2B-1)\right) ^{2}\right) $.

\section{\textbf{Integral kernel of $\widetilde{\wp }_{R}^{B}$ \ on the Bergman
space $\mathcal{A}^{B}(\mathbb{D})$}}

The CS transform associated with $\left\vert \widetilde{\kappa }%
_{z,B}\right\rangle $ is the isometric isomorphism $W_{B}:L^{2}(\mathbb{R}%
_{+})\rightarrow \mathcal{A}^{B}(\mathbb{D})$ defined by $\varphi \mapsto
W_{B}[\varphi ](z):=(1-z\bar{z})^{-B}\langle \varphi \left\vert \widetilde{%
\kappa }_{z,B}\right\rangle _{L^{2}\left( \mathbb{R}_{+}\right) }$, which
may also be called the second Bargmann transform \cite{IWGM}.
Explicitly, 
\begin{equation}
\label{r5.1}
W_{B}[f](z)=\sqrt{\frac{2}{\Gamma (2B)}}(1-z)^{-2B}\int_{0}^{+\infty }\xi
^{2B-\frac{1}{2}}\exp \left( -\frac{1}{2}\left( \frac{1+z}{1-z}\right) \xi
^{2}\right) d\xi ,\ z\in \mathbb{D}.  
\end{equation}
Now, to transfer the operator $\wp _{R}^{B}$ to be acting on functions $f$ $%
\in $ $\mathcal{A}^{B}(\mathbb{D}),$ we use the relation $W_{B}\circ \wp
_{R}^{B}\circ W_{B}^{-1}\equiv \widetilde{\wp }_{R}^{B}$. \ We, precisely,
obtain that 
\begin{equation}
\label{r5.2}
\widetilde{\wp }_{R}^{B}[f](w)=\int\limits_{\mathbb{D}}P_{R}^{B}(z,w)f(z)(1-z%
\bar{z})^{2B-2}d\eta (z)  
\end{equation}
where the integral kernel is given by 
\begin{equation}
\label{r5.3}
P_{R}^{B}(z,w)=\frac{(2B-1)^{2}R}{(1-R\bar{z}w)^{2B}}F_{1}\left(
2-2B,1-2B,2B,2,R,\frac{R-R\bar{z}w}{1-R\bar{z}w}\right)  
\end{equation}
In particular, at the limit $R\rightarrow 1$, 
\begin{equation}
\label{r5.4}
\lim_{R\rightarrow 1}P_{R}^{B}(z,w)=\frac{(2B-1)}{\left( 1-\bar{z}w\right)
^{2B}},  
\end{equation}
which is the reproducing kernel of the Bergman space $\mathcal{A}^{B}(%
\mathbb{D}).$ Here, $F_{1}$ denotes the Appel hypergeometric double series%
\begin{equation*}
F_{1}\left( \alpha ,\beta ,\gamma ;\omega ;u,v\right)
=\sum\limits_{j,k=0}^{\infty }\frac{\left( \alpha \right) _{j+k}\left( \beta
\right) _{j}\left( \gamma \right) _{k}}{j!k!\left( \omega \right) _{j+k}}%
u^{j}v^{k},\left\vert u\right\vert <1,\left\vert v\right\vert <1.
\end{equation*}

\bigskip \smallskip

To prove $\eqref{r5.2} ,$ let us take $f\in \mathcal{A}^{B}(\mathbb{D})$
and apply the inverse of $W_{B}$ as 
\begin{equation}
\label{r5.5}
W_{B}^{-1}[f](\xi )=\int\limits_{\mathbb{D}}f(z)\left\langle \xi \right\vert 
\widetilde{\kappa }_{z,B}\rangle (1-z\bar{z})^{-B}d\eta _{B}(z),\text{ }\xi
>0.  
\end{equation}
Next, we proceed by the action of $\wp _{R}^{B}$ on $W_{B}^{-1}[f],$ which
successively gives

\begin{equation}
\label{r5.6}
\wp _{R}^{B}\left[ W_{B}^{-1}[f]\right] (y)=\sum_{k=0}^{+\infty }\lambda
_{k}^{B,R}\left\langle \ell _{k}^{B}|\int\limits_{\mathbb{D}}f(z)|\widetilde{%
\kappa }_{z,B}\rangle (1-z\bar{z})^{-B}d\eta _{B}(z)\right\rangle
\left\langle y|\ell _{k}^{B}\right\rangle  
\end{equation}

\begin{equation}
\label{r5.7}
=\sum_{k=0}^{+\infty }\lambda _{k}^{B,R}\left( \int_{\mathbb{R}_{+}}%
\overline{\ell _{k}^{B}(\xi )}\left( \int_{\mathbb{D}}f(z)\left\langle
x\right\vert \widetilde{\kappa }_{z,B}\rangle (1-z\bar{z})^{-B}d\eta
_{B}(z)\right) d\xi \right) \ell _{k}^{B}(y)  
\end{equation}

\begin{equation}
\label{r5.8}
=\sum_{k=0}^{+\infty }\lambda _{k}^{B,R}\int_{\mathbb{D}}f(z)\left( (1-z\bar{%
z})^{-B}\overline{\int_{\mathbb{R}_{+}}\ell _{k}^{B}(\xi )\overline{%
\left\langle \xi \right\vert \widetilde{\kappa }_{z,B}\rangle }}dx\right)
d\eta _{B}\left( z\right) \ell _{k}^{B}(y).  
\end{equation}
Note that the last equation may be rewritten as
\begin{equation}
\label{r5.9}
\wp _{R}^{B}\left[ W_{B}^{-1}[f]\right] (y)=\sum_{k=0}^{+\infty }\lambda
_{k}^{B,R}\left[ \int_{\mathbb{D}}f(z)\overline{W_{B}[\ell _{k}^{B}](z)}%
d\eta _{B}(z)\right] \ell _{k}^{B}(y).  
\end{equation}
We again apply $W_{B}$ to $\eqref{r5.9}:$ 
\begin{equation}
\label{r5.10}
W_{B}\left[ \wp _{R}^{B}\left[ W_{B}^{-1}[f]\right] \right]
(w)=\sum_{k=0}^{+\infty }\lambda _{k}^{B,R}\left[ \int_{\mathbb{D}}f(z)%
\overline{W_{B}[\ell _{k}^{B}](z)}d\eta _{B}(z)\right] W_{B}[\ell
_{k}^{B}](w).  
\end{equation}
Since $W_{B}[\ell _{k}^{B}](w)=C_{k}^{B}(w),$ then we obtain
\begin{equation}
\label{r5.11}
\widetilde{\wp }_{R}^{B}[f](w)=\int_{\mathbb{D}}\left[ \sum_{k=0}^{+\infty
}\lambda _{k}^{B,R}\overline{C_{k}^{B}(z)}C_{k}^{B}(w)\right] d\eta _{B}(z).
\end{equation}
Hence, the kernel integral function is given by 
\begin{equation}
\label{r5.12}
P_{R}^{B}(z,w)=\sum_{k=0}^{+\infty }\lambda _{k}^{B,R}\overline{C_{k}^{B}(z)}%
C_{k}^{B}(w).  
\end{equation}
To write this kernel in a closed form, we recall $\eqref{r34} $ and $\eqref{r20} $, then $\eqref{r5.12} $ becomes%
\begin{equation}
\label{r5.13}
P_{R}^{B}(z,w)=(2B-1)\sum_{k=0}^{+\infty }\left[ \frac{1}{B(k+1,2B-1)}%
\int_{0}^{R}\rho ^{k}(1-\rho )^{2B-2}d\rho \right] \frac{\Gamma (2B+k)}{%
k!\Gamma (2B)}\bar{z}^{k}w^{k}  
\end{equation}
\begin{equation}
\label{r5.14}
=(2B-1)^{2}\sum_{k=0}^{+\infty }\frac{\Gamma (2B+k)\Gamma (2B+k)}{\Gamma
(2B)\Gamma (2B)}\frac{1}{k!}\frac{(\bar{z}w)^{k}}{k!}\left(
\int_{0}^{R}t^{k}(1-t)^{2B-2}dt\right)  
\end{equation}
\begin{equation}
\label{r5.15}
=(2B-1)^{2}\int_{0}^{R}(1-t)^{2B-2}\left( \sum_{k=0}^{+\infty }\frac{%
(2B)_{k}(2B)_{k}}{(1)_{k}}\frac{(t\bar{z}w)^{k}}{k!}\right) dt  
\end{equation}
The sum inside the integral can be presented as the Gauss hypergeometric
function ${}_{2}F_{1}$ as 
\begin{equation}
\label{r5.16}
P_{R}^{B}(z,w)=(2B-1)^{2}\int_{0}^{R}(1-t)^{2B-2}{}_{2}F_{1}(2B,2B,1,t\bar{z}%
w)dt. 
\end{equation}
By setting $2B=\alpha $, $\bar{z}w=\omega $ and 
\begin{equation}
\label{r5.17}
I_{R}=\int_{0}^{R}{}_{2}F_{1}(\alpha ,\alpha ,1;\omega t)(1-t)^{\alpha -2}dt.
\end{equation}
By making use of the formula   \cite[p.316]{PBM}  
: 
\begin{equation}
\label{r5.18}
\int_{0}^{y}x^{c-1}(y-x)^{\beta -1}(1-xz)^{-\tau }{}_{2}F_{1}(a,b,c;wx)dx=%
\mathcal{B}(c,\beta )y^{c+\beta -1}(1-yz)^{-\tau }F_{3}(\tau ,a,\beta
,b,c+\beta ,\frac{yz}{yz-1};wy),  
\end{equation}
$y,Re(c),Re(\beta )>0;|arg(1-wy)|,|arg(1-z)|<\pi $, for parameters $%
y=R,x=t,c=1,\beta =1,z=1,\tau =2-2B,a=2B,b=2B$, the integral $\eqref{r5.17} $ takes the form 
\begin{equation}
\label{r5.19}
I_{R}=R(1-R)^{\alpha -2}F_{3}(2-\alpha ,\alpha ,1,\alpha ,2;\frac{R}{R-1}%
,\omega R)  
\end{equation}
Next, we may apply the transformation \cite[p.450]{PBM}: 
\begin{equation}
\label{r5.20}
F_{3}\left( a,a^{\prime },b,b^{\prime };a+a^{\prime },w,z\right)
=(1-z)^{-b^{\prime }}F_{1}\left( a,b,b^{\prime };a+a^{\prime };w,\frac{z}{z-1%
}\right)  
\end{equation}
to rewrite the $F_{3}$ hypergeometric function in $\eqref{r5.19} $ as 
\begin{equation}
\label{r5.21}
F_{3}\left( 2-\alpha ,\alpha ,1,\alpha ,\frac{R}{R-1},\omega R\right)
=(1-\omega R)^{-\alpha }F_{1}\left( 2-\alpha ,1,\alpha ,2,\frac{R}{R-1},%
\frac{R\omega }{R\omega -1}\right) .  
\end{equation}%
Therefore, Eq.$\eqref{r5.19} $ reads 
\begin{equation}
\label{r5.22}
I_{R}=R(1-R)^{\alpha -2}(1-\omega R)^{-\alpha }F_{1}\left( 2-\alpha
,1,\alpha ,2,\frac{R}{R-1},\frac{R\omega }{R\omega -1}\right) . 
\end{equation}
By applying the transformation 
\begin{equation}
\label{r5.23}
F_{1}\left( a,b_{1},b_{2},c;X;Y\right)
=(1-X)^{-b_{1}}(1-Y)^{-b_{2}}F_{1}\left( c-a,b_{1},b_{2};c,\frac{X}{X-1},%
\frac{Y}{Y-1}\right)
\end{equation}
we may reduce $I_{R}$ as 
\begin{equation}
\label{r5.24}
I_{R}=R(1-R)^{\alpha -1}F_{1}\left( \alpha ,1,\alpha ,2;R,R\omega \right) . 
\end{equation}
Next, by using the symmetry relation 
\begin{equation}
\label{r5.25}
F_{1}\left( a,b,b^{\prime };c,z,u\right) =F_{1}\left( a,b^{\prime
},b;c,u,z\right)  
\end{equation}
together with the identity \cite[p.449]{PBM}
\begin{equation}
\label{r5.26}
F_{1}\left( a,b,b^{\prime };c,u,z\right) =(1-u)^{c-a-b}(1-z)^{-b^{\prime
}}F_{1}\left( c-a,c-b-b^{\prime },b^{\prime },c,u,\frac{u-z}{1-z}\right) 
\end{equation}
enable us to rewrite $I_{R}$ as 
\begin{equation}
\label{r5.27}
I_{R}=\frac{R}{(1-R\omega )^{2B}}F_{1}\left( 2-2B,1-2B,2B,2,R,\frac{%
R-R\omega }{1-R\omega }\right)  
\end{equation}
Therefore, the kernel function $\eqref{r5.16} $ reads 
\begin{equation}
\label{r5.28}
P_{R}^{B}(z,w)=\frac{(2B-1)^{2}R}{(1-R\bar{z}w)^{2B}}F_{1}\left(
2-2B,1-2B,2B,2,R,\frac{R-R\bar{z}w}{1-R\bar{z}w}\right) . 
\end{equation}
To check the limit of this kernel as $R\rightarrow 1$, we first observe that 
\begin{equation}
\label{r5.29}
F_{1}\left( 2-2B,1-2B,2B,2;R,\frac{R-R\bar{z}w}{1-R\bar{z}w}\right)
\rightarrow F_{1}\left( 2-2B,1-2B,2B,2,1,1\right) \text{ as \ }R\rightarrow
1.  
\end{equation}
By using the identity \cite[p.452]{PBM} 
\begin{equation}
\label{r5.30}
F_{1}\left( a,b,b^{\prime },c;Z;Z\right) ={}_{2}F_{1}\left( a,b+b^{\prime
},c;Z\right) ,  
\end{equation}
the obtained expression limit in $\eqref{r5.29}$ reduces as 
\begin{equation}
\label{r5.31}
F_{1}\left( 2-2B,1-2B,2B,2B,2;1,1\right) ={}_{2}F_{1}\left(
2-2B,1,2;1\right) .  
\end{equation}
Finally, we use the Gauss theorem \cite[p.489]{PBM}: 
\begin{equation}
\label{r5.32}
{}_{2}F_{1}(a,b,c;1)=\frac{\Gamma (c)\Gamma (c-a-b)}{\Gamma (c-a)\Gamma (c-b)%
}, \quad \rm{Re}(c-a-b)>0  
\end{equation}
for parameters $a=2-2B,b=1$ and $c=2$ to get that 
\begin{equation}
\label{r5.33}
{}_{2}F_{1}\left( 2-2B,1,2;1\right) =\frac{1}{2B-1}.  
\end{equation}
This, leads to the limit 
\begin{equation}
\label{r5.34}
\lim_{R\rightarrow 1}P_{R}^{B}(z,w)=\frac{(2B-1)}{(1-\bar{z}w)^{2B}} 
\end{equation}
This completes the proof.

\smallskip

We end this section by observing that $\eqref{r5.8} $ may provides us
with a family of Hilbert spaces $\left( \text{RKHS}\right) $ indexed by the
continuous parameter $R\in \left] 0,1\right[ ,$ where each one would have $%
P_{R}^{B}(z,w)$ as its reproducing kernel for which the Eq. $\eqref{r5.12} $ will represent a Zaremba expansion \cite{SFH}. These
RKHS are natural generalizations of the Bergman space $\mathcal{A}^{B}(%
\mathbb{D})$ and deserve to be investigated in details in a futur work.


\begin{thebibliography}{99}

\bibitem{W} \bigskip M.W. Wong, Wavelet Transform and Localisation
Operators, Operator Theory: Advances and Applications, Vol. 136, (2002) Birkh%
\"{a}user, Basel

\bibitem{V} V.V. Dodonov, 'Nonclassical` states in quantum optics: a
squeezed review of the first 75 years, J.Opt. B: Quantum Semiclass. Opt. 4
(2002) R1-R33.

\bibitem{G} J.P. Gazeau, Coherent states in quantum physics, WILEY-VCH
Verlag GMBH \& Co. KGaA Weinheim (2009)

\bibitem{TP} S. T. Ali, J. P. Antoine and J. P. Gazeau, Coherent States,
Wavelets and Their Generalizations, second edition, Springer
Science+Business Media New York (2014)

\bibitem{RS} M. Reed and B. Simon, Methods of Modern Mathematical Physics,
Vols. I--IV \symbol{126}Academic, New York (1978)

\bibitem{Daub} I. Daubechies, Time-frequency localization operators: a
geometric phase space approach, \textit{IEEE transactions on information
theory}, Vol 34, N4 (1998)

\bibitem{PD1} D. Popov, Barut-Girardello coherent states for the
pseudo-harmonic oscillator, J. Phys. A: Math. Gen. 34, 5283-5296 \ (2001)

\bibitem{PD2} D. Popov, Gazeau-Klauder quasi-coherent states for the Morse
oscillator, Phys. Lett. A. vol 316 (6) "69-381 (2003)

\bibitem{Mou1} Z. Mouayn, Husimi's Q-function of the isotonic oscillator in a
generalized Binomial states representation, \textit{Math. Phys. Anal. Geom}.
17(3-4):289-303, (2014)

\bibitem{GJT} Ts Gantsog, A. Joshi and R. Tanas, Quantum Opt. vol 6 , 517-526
(1994$)$

\bibitem{FS} H-C. Fu and R. Sasaki, Negative binomial states of quantized
Radiation fields, arXiv:quant-ph/9610024v1

\bibitem{DMo} Duflo M. and Moore C.C., On the regular representation of a
Nonunimodular locally compact group, \textit{J. Funct. Anal,} \textbf{21},
2, pp.209-243 (1976)

\bibitem{AK} Aslaken E W and Klauder J R , J.Math.Phys 10, p.2267 (1969).

\bibitem{Mou2} Z. Mouayn, Characterization of hyperbolic Landau states by
coherent state transform, J. Phys. A: Math \& Gen vol 36 (29) 8071 (2003$)$

\bibitem{MaOb} {W. Magnus, F. Oberhettinger, R.P. Soni}, \textit{\ Formulas
and Theorems for the special Functions of Mathematical Physics}, 3rd edn.
Die Grundlehren der mathematischen Wissenschaften, vol. 52. Springer, New
York(1966)

\bibitem{BMMB} M. G. Benedict and B. Molnar, An algebraic construction of the
coherent states of the Morse potential based on supersymmetric quantum
mechanics,\textit{\ Phys. Rev}. 60 R 1737 (1999)

\bibitem{MP} P. M. Morse, Diatomic molecules according to the wave
mechanics.II. Vibrational levels, Phys. Rev. D. 57 (1929)

\bibitem{NMSL} M. M. Nieto and L. M. Simmons Jr: Coherent states for general
potentials. I Formalism, II. Confining one-dimensional examples; III Non
confining one-diemensional examples, Phys. Rev. D. Vol 20, 1321,1332,1342 \
(1979)

\bibitem{CDM} H. Chhaiba, N. Demni and Z. Mouayn, Analysis of generalized
negative binomial distributions attached to hyperbolic Landau levels, \emph{%
J. Math. Phys},\textbf{\ 57}, 072103 (2016)

\bibitem{IWGM} A. Intissar, F. El Wassouli, A. Ghanmi and Z. Mouayn.
Generalized second bargmann transforms associated with the hyperbolic landau
levels on the poincar disk. Annales Henri Poincar, 13 (4):513-524, 2012.

\bibitem{PBM} A. P. Prudnikov, Yu. A. Brychkov and O. I. Marichev. Integrals
and Series-More special functions. Volume \textbf{3}. 1986.

\bibitem{SFH} F. H. Szafraniec, Reproducing kernel propoerty and its space :
The basics , in D. Alpay (eds) Operator Theory, Springer, Basel (2015)
\end{thebibliography}
\end{document}